\documentclass[aps,prl,reprint,groupedaddress]{revtex4-1}
\usepackage{graphicx, color, soul}
\usepackage{siunitx}
\usepackage{amsmath}
\usepackage{amsfonts}
\usepackage{bm}
\usepackage{color,soul,url}

\newcommand{\Z}{{\mathbb{Z}}}
\newcommand{\bR}{{\bf R}}
\newcommand{\bG}{{\bf G}}
\newcommand{\vF}{{\rm v}_D}

\begin{document}
\title{Landau levels in strained two-dimensional photonic crystals}

\author{J. Guglielmon$^{1}$ and M. C. Rechtsman$^{1}$}
\affiliation{$^{1}$Department of Physics, The Pennsylvania State University, University Park, PA 16802, USA}

\author{M. I. Weinstein$^{2}$}
\affiliation{$^{2}$Department of Applied Physics and Applied Mathematics and Department of Mathematics, Columbia University, New York, NY, 10027 USA}

\begin{abstract}
The principal use of photonic crystals is to engineer the photonic density of states, which controls light-matter coupling. We theoretically show that strained 2D photonic crystals can generate artificial electromagnetic fields and highly degenerate Landau levels. Since photonic crystals are not described by tight-binding, we employ a multiscale expansion of the full wave equation. Using numerical simulations, we observe dispersive Landau levels which we show can be flattened by engineering a pseudoelectric field. Artificial fields yield a design principle for aperiodic nanophotonic systems.
\end{abstract}

\maketitle
In the presence of strain, graphene exhibits a remarkable effect: an inhomogeneous deformation of the lattice induces a strong pseudomagnetic field governing the low energy theory \cite{Kane1997,Guinea2009,Levy2010strain,Gomes:2012vy,Manes2013,Amorim2016}. If the strain is designed to produce a uniform pseudomagnetic field, highly degenerate Landau levels form at energies near the Dirac points. In photonic systems, methods for generating pseudomagnetic fields \cite{Umucalilar2011,Hafezi2011,Fang2012,Fang2017generalized} are of significant interest since photons are fundamentally uncharged and therefore do not directly respond to real magnetic fields. If Landau levels could be realized in the nanophotonic domain using strained photonic crystals, the associated large density of states could used to enhance light-matter interactions (e.g., the Purcell effect \cite{purcell1946spontaneous} or nonlinear phenomena \cite{krauss2008}).

Strain-induced pseudomagnetic fields have been demonstrated experimentally in photonic systems of coupled waveguide arrays \cite{rechtsman2012strain}, exciton-polariton condensates based on coupled cavities \cite{jamadi2020direct,Salerno2015}, and microwave systems of coupled resonators \cite{bellec2020observation}. Photonic Landau levels have also been discussed in the context of lasing models \cite{Schomerus2013} and strain and related ideas have been explored as a means for producing pseudomagnetism in acoustic systems \cite{Brendel2016Pseudomagnetic,Abbaszadeh2017,Wen2019}.
In the waveguide arrays studied in \cite{rechtsman2012strain}, however, time is mapped to a spatial dimension, and thus energy eigenvalues do not correspond to mode frequencies but to propagation constants. Hence, the Landau levels will not directly alter the photonic density of states. Demonstrations based on coupled resonators can be treated with the standard tight-binding framework often used to study strained graphene. Photonic crystals, however, are governed by the continuum Maxwell equations to which tight-binding models do not generally apply \cite{joannopoulos2011photonic}. 

In this Letter, we address the question: Since Dirac points generically emerge in the presence of certain symmetries, \cite{fefferman2012honeycomb,fefferman2014wave,lee2019elliptic} and therefore occur also in photonic crystals, can strain be used to generate pseudomagnetic fields for light in the nanophotonic domain? To answer this question, we use a two-scale expansion of solutions to the full continuum wave equation to show that pseudomagnetic and pseudoelectric fields are present in a class of deformed 2D photonic crystals. Our results, which apply to wave equations in non-dissipative media, require only that the strain be slowly varying and that the unstrained periodic structure exhibit Dirac points associated with a certain set of symmetries. The effective equations contain no free parameters. We make no assumptions about the magnitude of the material (index) contrast and our results do not require  an effective tight-binding model.

We assess the validity of our effective theory by performing full-wave numerical simulations in an experimentally realistic strained photonic crystal of air holes embedded in silicon. 
These simulations demonstrate high density of states at energies corresponding to the Landau levels of the effective theory. However, the Landau levels are weakly dispersive and we find that producing nearly flat (non-dispersive) Landau levels in such a photonic crystal can be achieved  using a new ingredient: a strain that, on the level of the effective equations, generates a pseudoelectric field (in addition to the pseudomagnetic field), and on the level of the full wave equation, acts to flatten the bands.

We begin by considering light propagating in the plane of a two-dimensional photonic crystal: a medium consisting of a real, spatially varying scalar dielectric $\varepsilon(\mathbf{x})$ that is uniform in the $x_3$ direction, so that we may take $\mathbf{x}=[x_1,x_2]$. The solutions can be classified as having either TE or TM polarization. For time-harmonic solutions with electric and magnetic fields $\mathbf{E}(\mathbf{x},t) = \mathbf{E}(\mathbf{x})e^{-i \omega t}$ and  $\mathbf{H}(\mathbf{x},t) = \mathbf{H}(\mathbf{x})e^{-i \omega t}$, the modes are governed by the scalar  Helmholtz equation 
\begin{equation}\label{eqn_maxwell}
-\nabla \cdot \big(\xi(\mathbf{x}) \nabla \big) \psi(\mathbf{x}) = (\omega/c)^2 \rho(\mathbf{x})\psi(\mathbf{x}),
\end{equation}
where $c$ denotes the vacuum speed of light. For TE polarization, $\xi(\mathbf{x}) = 1/\varepsilon(\mathbf{x})$, $\rho(\mathbf{x}) = 1$, and the scalar function $\psi(\mathbf{x})$ gives the magnetic field $\mathbf{H}(\mathbf{x}) = \psi(\mathbf{x}) \hat{\mathbf{z}}$. For TM polarization, $\xi(\mathbf{x}) = 1$, $\rho(\mathbf{x}) = \varepsilon(\mathbf{x})$, and $\psi(\mathbf{x})$ gives the electric field $\mathbf{E}(\mathbf{x}) = \psi(\mathbf{x})\hat{\mathbf{z}}$.

To obtain a structure possessing Dirac points, we require that $\varepsilon(\mathbf{x})$ is inversion symmetric, $C_3$ rotation invariant, and translation invariant with respect to the triangular lattice $\Z\bR_1\oplus\Z\bR_2$, where $\bR_1=a[1,0]$ and $\bR_2=a[1/2,\sqrt{3}/2]$ and $a$ denotes the lattice constant. It was proved in \cite{fefferman2012honeycomb,fefferman2014wave,lee2019elliptic} for continuum media that these conditions imply the existence of Dirac points at the Brillouin zone vertices $\mathbf{K}=(\bG_1-\bG_2)/3$,  $\mathbf{K}' = -\mathbf{K}$, where $\bG_1$ and $\bG_2$ are the reciprocal lattice vectors. For a Dirac point occurring at quasimomentum $\mathbf{k}_D$ and energy $E_D$, two consecutive bands $E_\pm(\mathbf{k})$ of the operator defined in Eq. \eqref{eqn_maxwell}  (with $E=\omega^2/c^2$) exhibit a conical intersection 
\begin{equation}  
E_\pm(\mathbf{k})\ =\ E_D\ \pm\ \vF\ |\mathbf{k}-\mathbf{k}_D|\ \left(1+\mathcal{O}(|\mathbf{k}-\mathbf{k}_D|\right)\ \label{cones}
\end{equation}
for $\mathbf{k}$ near $\mathbf{k}_D$.

Since $\omega^2=c^2 E$ there are positive and negative  branches of frequencies, $\omega$, obtained from \eqref{cones}. We focus on the positive branch, yielding
 \begin{equation}
 \omega_\pm(\mathbf{k})=\omega_D\pm (\vF c^2/(2\omega_D))\ |\mathbf{k}-\mathbf{k}_D|\ \left(1+\mathcal{O}(|\mathbf{k}-\mathbf{k}_D|\right).
 \label{om-cone}
 \end{equation}

We now focus on the Dirac point at $\mathbf{K}$; the case  
$\mathbf{K}' = -\mathbf{K}$ can be treated similarly (with a pseudomagnetic field that points in the opposite direction). We denote the two energy-degenerate states at the Dirac point by $\Phi_1(\mathbf{x}), \Phi_2(\mathbf{x})$. 
These can be taken to satisfy $\Phi_2(\mathbf{x})=\overline{\Phi_1(-\mathbf{x})}$, $\mathcal {R}[\Phi_1]=e^{2\pi i/3}\Phi_1$, and $\mathcal {R}[\Phi_2]=e^{-2\pi i/3}\Phi_2$, where $\mathcal {R}[f](\mathbf{x})= f(R^\dagger\mathbf{x})$ and $R$ is a $2\times2$ matrix of rotation by $2\pi/3$. We use a normalization $\langle \Phi_i|\Phi_j\rangle_\rho = \delta_{ij}$. See the Supplemental Material for more details and definitions of the two inner products $\langle f|g\rangle$ and $\langle f|g\rangle_\rho$ used in the text.
  
There are two parameters in the effective theory, which are computed from the eigenmodes, $\Phi_i(\mathbf{x})$, of the unstrained system, and which determine the behavior of the strained system:
\begin{align}\label{eqn_vf}
\vF &= \big\langle \Phi_1\big| \big\{\hat{p}_1,\hat{\xi}\big\} \big| \Phi_2\big\rangle\\
\label{eqn_bstar}
b_\star &= \big\langle \Phi_1 \big|\hat{p}_1 \hat{\xi}\, \hat{p}_1 \big| \Phi_2\big\rangle\ .
\end{align}
The {\it Dirac velocity}, $\vF$, is  associated with the Dirac point of the periodic structure
 and $b_\star$ emerges in connection with the induced pseudomagnetic field; see the Supplemental Materials.
Here, $\hat{p}_1=-i\partial_{x_1}$, $\big\{\hat{p}_1,\hat{\xi}\big\} =\hat{p}_1\hat{\xi} + \hat{\xi}\,\hat{p}_1$, and $\hat{\xi}$ is the operator that acts in position space via multiplication by $\xi(\mathbf{x})$. In the Supplemental Material, we show that both $\vF$ and $b_\star$ can be made to be real and positive by an appropriate choice of a coordinate system and a phase convention for the eigenstates. We assume these choices have been made.

\begin{figure}
\centering
\includegraphics[width = 0.9\columnwidth]{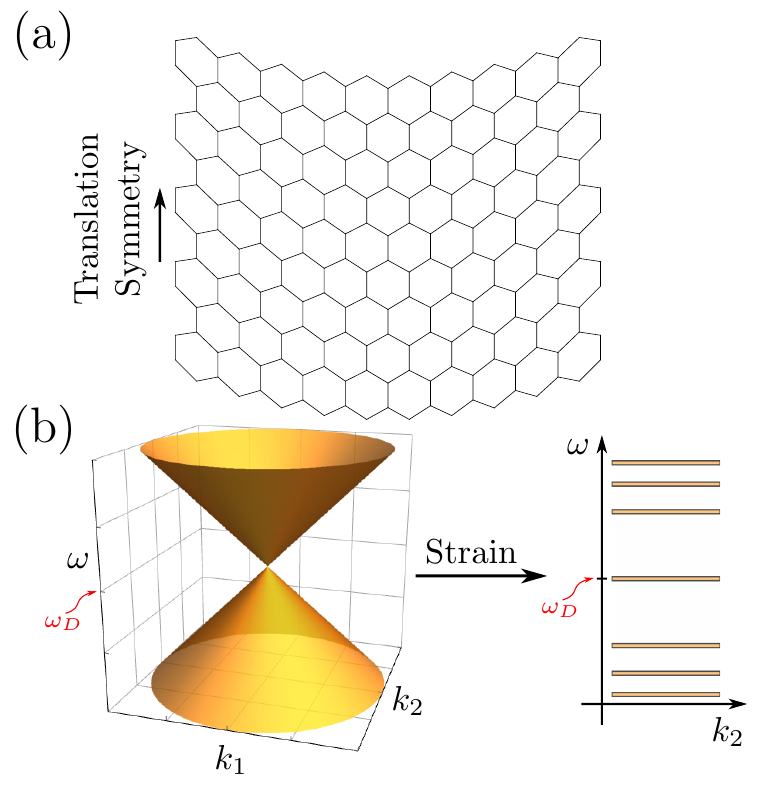}
\caption{\label{fig_schematic_strain}(a) Illustration of a strain that produces a Landau gauge vector potential for a uniform pseudomagnetic field. (b) Schematic illustration of the effect of the strain on the spectrum. In a neighborhood of $\omega_D$, the Dirac cone is transformed into a sequence of discrete  Landau levels.}
\end{figure}

\begin{figure*}
\centering
\includegraphics[width = 1\textwidth]{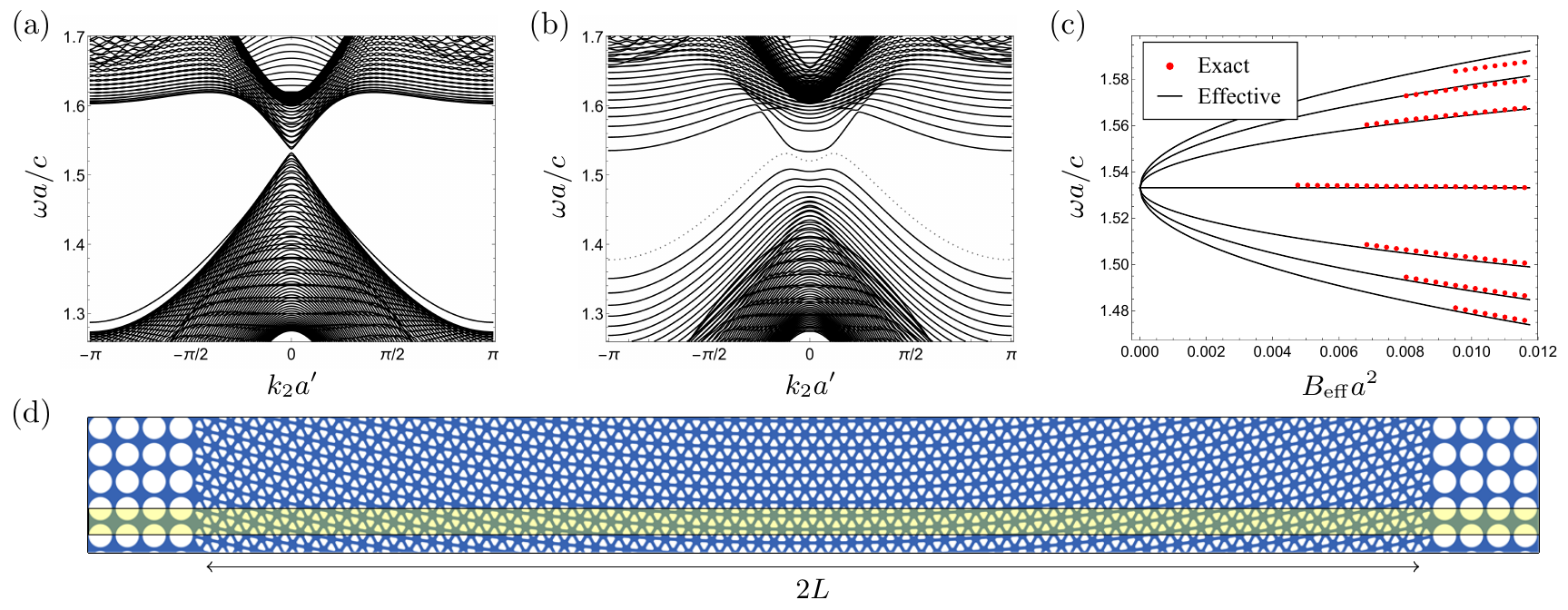}
\caption{\label{fig_strained_bands}(a) Full-wave band structure for TE polarized modes, showing the Dirac point of the unstrained honeycomb structure. (b) Band structure in the presence of strain, showing the emergence of Landau levels. The dashed curve corresponds a spurious mode supported at the computational boundary. (c) Level spacing as a function of the pseudomagnetic field strength. Solid curves show the prediction of Eq. (\ref{eqn_Heff}). Red points show the spacings obtained numerically from the (full-wave) band structures at $k_2=0$. (d) The strained dielectric used to generate panel (b), with air holes shown in white, silicon shown in blue, and the unit cell highlighted in yellow.}
\end{figure*}

A strained dielectric $\varepsilon'(\mathbf{x})$ is obtained by displacing each point $\mathbf{x}$ of the original dielectric  to a new location $T(\mathbf{x}) = \mathbf{x} + \mathbf{u}(\mathbf{x})$  giving $\varepsilon'(\mathbf{x} + \mathbf{u}(\mathbf{x})) = \varepsilon(\mathbf{x})$, where $\mathbf{u}(\mathbf{x})=\left(u_1(\mathbf{x}),u_2(\mathbf{x})\right)$. The corresponding strain matrix is denoted $U(\mathbf{x})=\frac12\Big(D_\mathbf{x}\mathbf{u}+\left(D_\mathbf{x}\mathbf{u}\right)^\top\Big)$ with entries 
$U_{ij} = \frac{1}{2} \left(\frac{\partial u_j}{\partial x_i} + \frac{\partial u_i}{\partial x_j}\right)$.
We assume that $\mathbf{u}(\mathbf{x})$ is deformed on a scale which is large compared 
with the lattice constant in the sense that $\mathbf{u}(\mathbf{x}) = \mathbf{f}(\kappa \mathbf{x})$, where $\kappa$ is a small parameter, with units $\textrm{length}^{-1}$, that measures the length scale over which  the deformation varies. Hence, $U({\bf x})\sim\mathcal{O}(\kappa)$ and one can think of $\kappa$ as measuring the strain strength.

Using a general systematic perturbation theory in the small parameter $\kappa$ (see Supplemental Material), we show  that the  strained dielectric $\varepsilon'(\mathbf{x})$ has modes with a two-scale spatial structure in which a pair of slowly varying amplitude functions $\alpha_i$, with ${i=1 \text{ or } 2}$, modulate the Dirac point eigenmodes:
\begin{align}
\psi(\mathbf{x}) &= \sum_{i=1}^2 \alpha_i\left(T^{-1}(\mathbf{x})\right) \Phi_i(T^{-1}(\mathbf{x})) + \mathcal{O}(\kappa)\label{two-scale}.
\end{align}
[As before, slowly varying is understood to mean $\alpha_i(\mathbf{x}) = g_i(\kappa \mathbf{x})$].
These modes have associated perturbed frequencies ${\omega = \omega_D \ +\ \frac{c^2}{2\omega_D}\ E_1\  +\ \mathcal{O}(\kappa^2)}$.
  The amplitude functions $\alpha_i(\mathbf{x})$ and frequency perturbation $E_1$  (of order $\kappa$) are determined by an eigenvalue problem $\mathcal{H}_{\rm eff} \alpha'(\mathbf{x}) = E_1 \alpha'(\mathbf{x})$,
where $\alpha'(\mathbf{x}) = [\alpha_2(\mathbf{x}),\ \alpha_1(\mathbf{x})]^T$ and $\mathcal{H}_{\rm eff} 
$ is a 2D Dirac Hamiltonian:
\begin{align}\label{eqn_Heff}
\mathcal{H}_{\rm eff} 
&= \vF [-i\nabla_\mathbf{x}-\mathbf{A}_{\rm eff}(\mathbf{x})]\cdot\mathbf{\sigma} + W_{\rm eff}(\mathbf{x})\ \sigma_0\ ,
\end{align}
where $\mathbf{\sigma} = (\sigma_1,\sigma_2)$ with $\sigma_j$ denoting the Pauli matrices. The effective magnetic vector potential $\mathbf{A}_{\rm eff}(\mathbf{x})$ and  electric potential $W_{\rm eff}(\mathbf{x})$ are given by
\begin{align}
\label{eqn_Aeff}
\mathbf{A}_{\rm eff}(\mathbf{x})
&=
\left(\frac{2b_\star}{\vF}\right)
\begin{bmatrix}
+{\scriptstyle}\text{tr}\big(U(\mathbf{x}) \sigma_3\big)\\
-{\scriptstyle}\text{tr}\big(U(\mathbf{x}) \sigma_1\big)
\end{bmatrix}\\
\label{eqn_Veff}
W_{\rm eff}(\mathbf{x}) &= -\left(\frac{\omega_D}{c}\right)^2 \text{tr}\big(U(\mathbf{x})\sigma_0\big)= -\left(\frac{\omega_D}{c}\right)^2\ \nabla\cdot\mathbf{u}(\mathbf{x}).
\end{align}
We emphasize that $\mathcal{H}_{\rm eff}$, $\mathbf{A}_{\rm eff}$ and $W_{\rm eff}$ emerge from a first principles derivation and depend on $\vF$ and $b_\star$ defined fully in terms of the Dirac point eigenmodes of the unstrained structure; see \eqref{eqn_vf}-\eqref{eqn_bstar}. In the Supplemental Materials, we show that the above quantities transform as expected under rotations.
   
We now focus on the case in which the strain produces a constant pseudomagnetic field perpendicular to the plane of the structure. As a concrete example, we consider a honeycomb lattice of air holes ($\varepsilon = 1$) embedded in silicon ($\varepsilon = 12.11$) operating in TE polarization. We take the air holes to have a triangular shape as in \cite{Barik2016two} to ensure that the frequency $\omega_D$ is not crossed by the same or other bands, {\it i.e.} the structure is semi-metallic at energy $(\omega_D/c)^2$; see  \cite{FLW-2d_edge:16, DW:19}.  To improve numerical convergence, we take the corners of the triangles to be rounded. We take the triangle radius (center to corner) to be $r=0.27a$ with a rounded corner radius of $r_c = 0.1a$.

We numerically compute the modes of this structure using a plane wave eigensolver (MPB) \cite{johnson2001numerical}. The system has a Dirac point at $\omega_D = 1.533ca^{-1}$, where the first and second TE bands touch. The quantities in Eqs. (\ref{eqn_vf}) and (\ref{eqn_bstar}) are given by $\vF= 0.684a^{-1}$and $b_\star = 0.502a^{-2}$. We apply a strain generated by $\mathbf{u}(\mathbf{x}) = a[0,(\kappa x_1)^2]$, where $\kappa$ determines the strength of the strain. This deformation is illustrated schematically in Fig. \ref{fig_schematic_strain}(a). Note that while the strain breaks translation symmetry in the $x_1$ direction, the structure remains symmetric under translation by a distance $a' = \sqrt{3}a$ along the $x_2$ direction. From Eq. (\ref{eqn_Aeff}), the vector potential is $\mathbf{A}_{\rm eff}(\mathbf{x})= -(4a\kappa b_\star/\vF) [0,\kappa x_1]$, which is a Landau gauge vector potential for a constant effective magnetic field $\mathbf{B}_{\rm eff}=  \nabla \times \mathbf{A}_{\rm eff}=- \kappa^2\ B_0  \hat{\mathbf{z}}\ \ \textrm{with}\ \ B_0 = 4a b_\star/\vF.$ Since ${\nabla \cdot \mathbf{u} = 0}$, the strain-induced pseudoelectric potential vanishes: $W_{\rm eff}(\mathbf{x})=0$.

For a constant pseudomagnetic field and zero pseudoelectric potential, the eigenvalues of the Hamiltonian (\ref{eqn_Heff}) are well-known to form a series of Landau levels consisting of discrete eigenvalues, $E_1$,  of infinite multiplicity implying eigenvalues, $\omega$,  of the Helmholtz equation \eqref{eqn_maxwell}: 
\begin{equation}\label{eqn_spacings}
\omega\ =\  \omega_D  \pm \frac{\vF c^2}{\sqrt{2}\omega_D}\sqrt{n|\mathbf{B}_{\rm eff}(\kappa)|}\\ +\ \mathcal{O}(\kappa^2),\ \ n=0,1,2,\dots,
\end{equation}
where $|\mathbf{B}_{\rm eff}(\kappa)|=B_0\ \kappa^2$; see Supplemental Materials.

We compare this prediction to full numerical simulations by directly solving  Eq. (\ref{eqn_maxwell}) for the eigenmodes of the strained structure, again using a numerical plane wave eigensolver. We impose Bloch boundary conditions in the $x_2$ direction and effectively apply exponentially decaying boundary conditions in the non-periodic $x_1$ direction by padding the boundaries with a structure that exhibits a TE band gap for  $\omega\approx\omega_D$. Since the strain preserves translation symmetry along $x_2$, the Bloch momentum $k_2$ remains a good momentum. Since the supercell used for this strain pattern is invariant under translations by $\sqrt{3}a$ (as opposed to $a$) along $x_2$, both Dirac points (from $\mathbf{K}$ and $\mathbf{K}'$) reside at $k_2=0$. The system size along the non-periodic direction is $L =39a$ [see Fig. \ref{fig_strained_bands}(d)].

\begin{figure}
	\centering
	\includegraphics[width = 1\columnwidth]{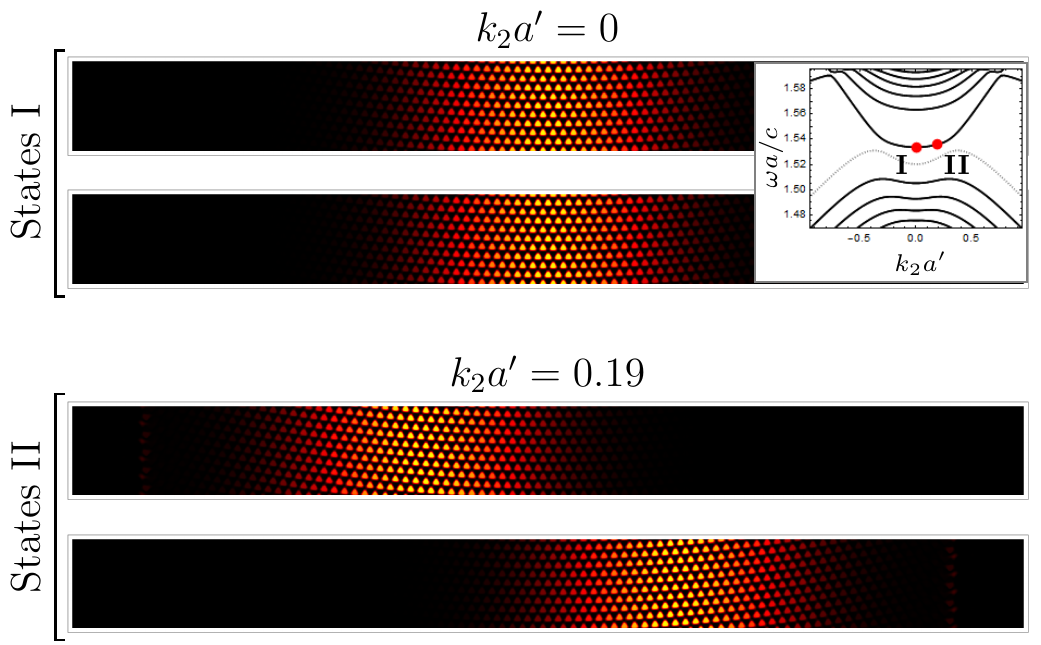}
	\caption{\label{fig_eigenstates}
		Numerically computed $0$-th Landau level eigenstates for the band structure of Fig. \ref{fig_strained_bands}(b). Eigenstates correspond to the two values of $k_2$  (I and II) shown in the zoomed inset. The modes come in two-fold degenerate pairs, are localized along the $x_1$ direction, and shift horizontally as $k_2$ is varied. 
	}
\end{figure}

The numerically computed band structures are shown in Fig. \ref{fig_strained_bands}. Upon applying the strain, the Dirac point of Fig. \ref{fig_strained_bands}(a) splits into a sequence of discrete Landau levels shown in Fig. \ref{fig_strained_bands}(b), which was obtained using $\kappa = 0.0548a^{-1}$. In Fig. \ref{fig_strained_bands}(c), we compare, as a function of strain strength, the level spacings predicted by Eq. (\ref{eqn_spacings}) to the numerically computed level spacings obtained from the band structures at $k_2=0$, with the results showing good agreement. Our multiscale analysis, which approximates states by spectral components near the Dirac point, is valid for $|k_2|\le C\kappa$, for some constant $C$ and all $\kappa a\ll1$. As the strain is reduced, the Landau level eigenstates become progressively more delocalized along the $x_1$ direction and eventually reach the boundary of the computational domain. Hence, the series of simulation points in Fig. \ref{fig_strained_bands}(c) is terminated at weak strain. 

\begin{figure}
	\centering
	\includegraphics[width = 0.9\columnwidth]{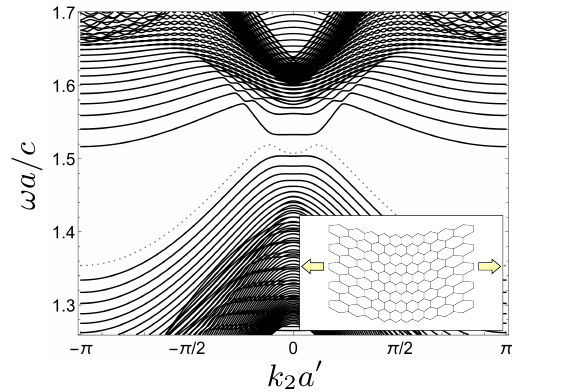}
	\caption{\label{fig_flattened} Band structure obtained by beginning with the strained structure associated with Fig. \ref{fig_strained_bands}(b) and applying an additional strain along the $x_1$ direction to flatten the Landau levels. Inset schematically illustrates the additional strain.}
\end{figure}

In Fig. \ref{fig_eigenstates}, we show representative numerically obtained eigenstates. The modes come in two-fold degenerate pairs since the supercell used in our calculation places both Dirac points at $k_2=0$. For a fixed Landau level, the eigenstate at a given $k_2$  is localized in the $x_1$ direction and centered about $\tilde{x}$, where $\tilde{x}$ varies with $k_2$. This is the expected behavior for the standard Landau gauge Hamiltonian for a particle in a uniform magnetic field (see Supplemental Material).
  
Although the effective theory predicts flat Landau levels, the levels in Fig. \ref{fig_strained_bands}(b) are weakly dispersive. This dispersion arises from contributions of order $\kappa^2$ which are neglected from the effective theory. In the Supplemental Material, we motivate the use of a deformation of the form $\mathbf{u}(\mathbf{x}) = a [\beta (\kappa x_1)^3\vspace{3pt},(\kappa x_1)^2]$ for mitigating this dispersion. On the level of the effective equations, this yields $\mathbf{B}(\mathbf{x}) = -(4a\kappa^2b_\star/\vF) \hat{\mathbf{z}}$ as before, but now with a quadratic potential $W_{\rm eff}(\mathbf{x}) = 3a\beta\kappa(\omega_D/c)^2  (\kappa x_1)^2$, corresponding to a pseudoelectric (as opposed to pseudomagnetic) field, which can be used as a tool to compensate for the dispersion. Note that this is not required in the graphene picture of pseudomagnetism~\cite{Guinea2009}. It is required in the continuum photonic crystal setting because of a lack of an accurate nearest-neighbor tight-binding model. The numerically computed band structures that result from taking $\kappa = 0.0548a^{-1}$ and $\beta=0.0380$ are shown in Fig. \ref{fig_flattened}, where we see a clear flattening of the Landau levels.

In conclusion, we have shown that, for a class of 2D photonic crystals possessing Dirac points, strain produces pseudoelectric and pseudomagnetic fields for photons. Explicit expressions for all parameters of the effective Hamiltonian are given in terms of the Bloch eigenmodes at the Dirac point of the unstrained structure. There are no free parameters. The modes of the strained structure are constructed as slow modulations of  deformed Dirac point eigenmodes. The modulations are governed by a Dirac equation with effective magnetic and electric potentials. Using a specific strain pattern, we have demonstrated the emergence of Landau levels in a photonic crystal that could be realized using standard fabrication techniques in silicon photonics. We found that a conventional strain (as in graphene) gives rise to dispersive Landau levels, but that dispersion can be corrected for (i.e., the bands can be flattened) using a strain that induces an additional pseudoelectric field that does not alter the original pseudomagnetic field.

Multiscale analysis enables treatment of dielectric structures that cannot be treated with band theory. This technique could be applied to general aperiodic media that arise as slowly varying deformations of periodic media, including media not amenable to tight-binding methods. This approach provides an analytical handle on and physical intuition for such systems that, in many settings, require prohibitive numerical simulations.  We envision that strain-induced pseudomagnetic and pseudoelectric fields will be useful for applications, particularly in the nanophotonic domain, such as chip-scale nonlinear optics and coupling to quantum emitters, where the high density-of-states associated with flat bands implies strong enhancement of light-matter interaction. 

\acknowledgements{M.C.R. and J.G. acknowledge the support of the National Science Foundation under award number DMS-1620422, as well as the Packard Foundation under fellowship number 2017-66821. M.I.W was supported in part by DMS-1412560, DMS-1620418, DMS-1908657 and Simons Foundation Math + X Investigator Award \#376319. He acknowledges stimulating discussions with A. Drouot and J. Shapiro.}
\bibliography{LandauLevelsRefsV18}
\end{document}